\documentclass[prl,twocolumn,showpacs]{revtex4}
\usepackage[dvips]{graphicx}
\usepackage[dvips]{color}
\usepackage{bm}

\begin{document}
\title{Anisotropic minimal conductivity of graphene bilayers}

\author{Ali G. Moghaddam and Malek Zareyan}

\affiliation{Institute for Advanced Studies in Basic Sciences,
P.O. Box 45195-1159, 45195 Zanjan, Iran}

\begin{abstract}
Fermi line of bilayer graphene at zero energy is transformed into
four separated points positioned trigonally at the corner of the
hexagonal first Brillouin zone. We show that as a result of this
trigonal splitting the minimal conductivity of an undoped bilayer
graphene strip becomes anisotropic with respect to the orientation
$\theta$ of the connected electrodes and finds a dependence on its
length $L$ on the characteristic scale $\ell=\pi/\Delta k\simeq 50
nm$ determined by the inverse of k-space distance of two Dirac
points. The minimum conductivity increases from a universal
isotropic value $\sigma^{min}_{\bot}=(8/\pi)e^2/h$ for a short
strip $L\ll \ell$ to a higher anisotropic value for longer strips,
which in the limit of $L\gg \ell$ varies from
$(7/3)\sigma^{min}_{\bot}$ at $\theta=0$ to $3\sigma^{min}_{\bot}$
over an angle range $\Delta \theta\sim \ell/L$.
\end{abstract}
\pacs{81.05.Uw, 73.23.-b, 73.23.Ad, 73.63.-b} \maketitle

The recent realization of isolated graphene
\cite{geim04,geim05,kim}, a two dimensional hexagonal lattice of
carbon atoms, and its bilayer \cite{geim06} has followed by
intensive studies which explored many intriguing properties of
these new carbon-based material \cite{castro-rmp}. They are
zero-gap semiconductors with their valance and conduction bands
touching each other at the corners of the hexagonal first
Brillouin zone, known as Dirac points. This specific band
structure in connection with pseudo-spin aspect which
characterizes the relative amplitude of electron wave function on
two different sublattices of the hexagonal structure, have given
the carriers a pseudo-relativistic chiral nature
\cite{geim05,geim06}. The chirality is believed to be origin of
most of peculiarities of quantum transport effects in single and
bilayer graphene \cite{cheianov-prbr,kats-natphys,castro-rmp}.
\par
One of the most important observations in graphene systems is the
existence of a nonzero minimal conductivity in the limit of
vanishing carrier density (at Dirac point) \cite{geim05}. This
effect which had been predicted theoretically long before the
experimental synthesis of graphene \cite{fradkin}, has been the
subject of several recent theoretical investigations and in most
studies a universal value $\sigma_0^{min}= (4/\pi) e^2/h$ for
minimum conductivity of monolayer graphene was found
\cite{ziegler,gusynin,peres,katsnelson1,mirlin,beenakker06}.
Although the theoretical value is $\pi$ times smaller than the
value measured in the early experiments, more recent experiments
have confirmed the predicted universal value $\sigma_0^{min}$ for
wide and short graphene strips \cite{lau,morpurgo}.

\par
\begin{figure}
\begin{centering}\vspace{-0.1cm}\par\end{centering}
\begin{centering}\includegraphics[width=3.9cm]{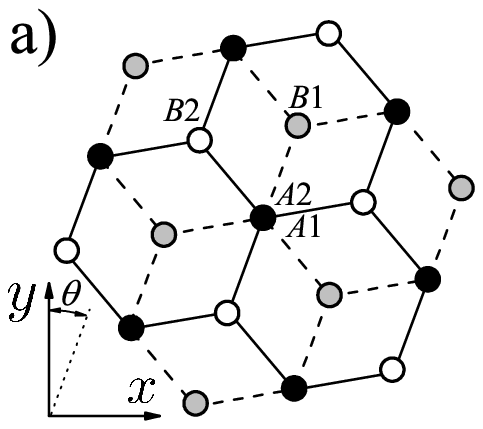}$\quad$\includegraphics[width=3.9cm]{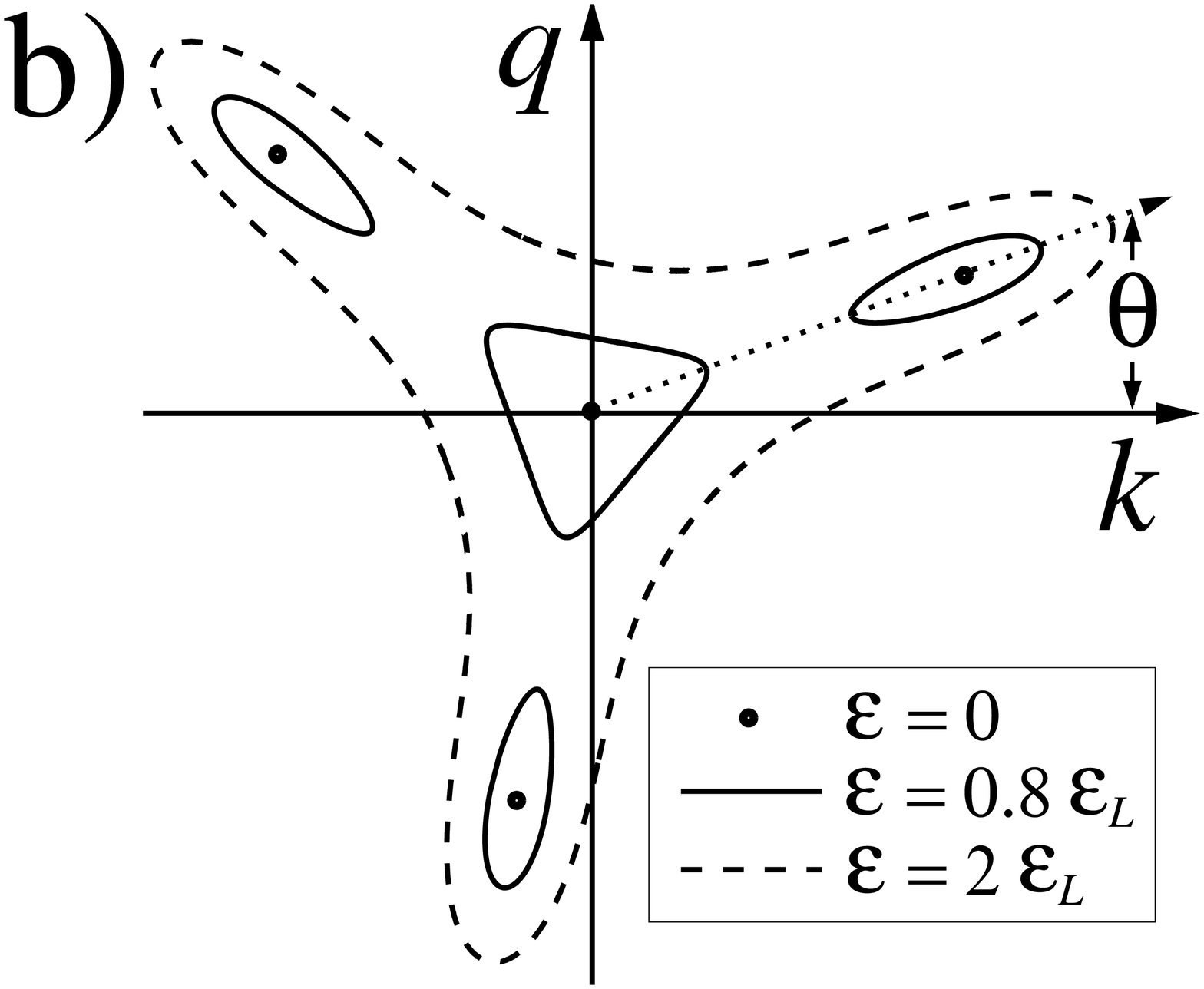}
\par\end{centering}
\caption{\label{fig1}{\small  (a) The bilayer lattice structure
projected onto the $x-y$ plane. Bonds between A1-B1 (A2-B2) in
bottom (top) layer are indicated by dashed (solid) lines; $\theta$
is the angle between the hexagonal lattice symmetry axis and
$y$-axis. (b) Constant-energy lines around the corner of the
hexagonal first Brillouin zone for different energies (measured in
units of Lifshitz transition energy $\varepsilon_{L}\approx 1
meV$) near the neutrality point. Trigonally warped Fermi line is
transformed into four separated points with one at the center and
three positioned trigonally around it for $\varepsilon=0$. }}
\end{figure}

For bilayer graphene the minimum conductivity is also measured to
be of order $e^2/h$ \cite{geim06,morozov}. Despite the similarity
of the chiral nature in single and bilayer graphene, the low
energy spectrum in bilayer is drastically different from the
linear dispersion of massless Dirac fermions in monolayer. The
spectrum in bilayer, which has a parabolic form at high energies,
acquires strong trigonal warping at low energies and undergoes the
so called Lifshitz transition at which the Fermi line is broken
into four separated pockets \cite{falko}. In the limit of zero
Fermi energy the pockets shrink into the points of which one is
located at the Dirac point and three others positioned around it
in a trigonal form (see Fig. \ref{fig1}(a)). The aim of the
present letter is to study effect of this Dirac point trigonal
splitting on the minimum conductivity of the bilayer, which will
also allow to distinguish between the effects of masslessness and
chirality. We employ the realistic model of a wide undoped bilayer
strip of length $L$ as the scattering region connecting two highly
doped regions as electrodes. Using a full Hamiltonian which takes
into account all intra and interlayer hoppings between nearest
neighbors atomic sites, we find that the effect of the trigonal
splitting in the minimum conductivity depends on $L$ as compared
to a characteristic length $\ell=\pi/\Delta k \simeq 50 nm$
determined by the inverse of the k-space separation $\Delta k$ of
two split Dirac points.

\par
For a short strip of $L\ll \ell$ the effect of trigonal splitting
is negligible and $\sigma^{min}_{\bot}= (8/\pi) e^2/h$, which is
twice the minimum conductivity of a monolayer showing that in this
limit the bilayer behaves as two independent single layer
connected in parallel. For finite length strip the minimal
conductivity increases above $\sigma^{min}_{\bot}$ and finds a
dependence on the angle $\theta$ between the orientation of the
electrodes and the hexagonal lattice symmetry axis. We find that
for a long strip  $L\gg \ell$ the anisotropic minimum conductivity
$\sigma^{min}(\theta)$ increases from $(7/3) \sigma^{min}_{\bot}$
at $\theta=0$ to $3\sigma^{min}_{\bot}$ over an angle range
$\Delta\theta \sim \ell/L$. Our results reveals importance of
trigonal splitting of the zero energy spectrum on the minimal
conductivity of bilayer graphene.

\par
To this end, there have been few theoretical investigations
devoted to the minimal conductivity in bilayer graphene
\cite{katsnelson2,cserti-prb,beenakker07,cserti-prl,ando,castro07}.
In Ref. \cite{beenakker07} a wide bilayer sheet with a constant
perpendicular interlayer hopping is considered to connect two
heavily doped electrode regions. This model, which ignores the
trigonal splitting, results in a minimum conductivity
$\sigma^{min}=\sigma^{min}_{\bot}$ \cite{beenakker07,cserti-prb},
establishing on the fact that for high energy electrons injected
from the metallic electrodes, the constant interlayer hopping does
not cause any significant effect and the bilayer sheets behaves as
two monolayer in parallel. On the other hand authors of Refs.
\cite{ando} and \cite{cserti-prl} have taken into account the
effect of strong trigonal warping within, respectively, Born
approximation and Kubo formula. They have found an isotropic and
constant minimal conductivity $(24/\pi) e^2/h$, which is three
times larger than $\sigma^{min}_{\bot}$ obtained in Refs.
\cite{beenakker07,cserti-prb}. However we note that the models
employed in Refs. \cite{cserti-prl,ando} do not include the effect
of electrodes which are present in a realistic experimental setup
for conductivity measurement. This could be in particular
important in graphene contacts due to the chirality of the
carriers and the resulting Klein tunneling phenomena
\cite{cheianov-prbr,kats-natphys}. Our study, taking into account
both of the trigonal splitting and the electrodes effect, reveals
anisotropy of the minimal conductivity and its dependence on
$L/\ell$ which also clarifies the origin of the disagreement
between the two above predictions.
\par

We consider a ballistic bilayer graphene sheet in $x-y$ plane
consisting of an undoped strip of length $L$ and width $W$ and two
heavily doped regions for $x < 0$ and $x > L$ on top of which bias
electrodes are deposited. The interfaces between electrode regions
and bilayer strip are oriented parallel to $y$ axis making an
angle $\theta$ with respect to the symmetry axis of the bilayer
lattice as indicated in Fig. \ref{fig1}(a).
\par
Bilayer graphene is made of two coupled single layer graphene with
different sites $A1, B1$ in the bottom layer and $A2, B2$ in the
top layer. The two layers are arranged according to Bernal
stacking in which every $A1$ site of bottom layer lies directly
below an $A2$ site in the top layer, as shown in Fig.
\ref{fig1}(a). Within the tight binding model of graphite
\cite{castro-rmp, castro07} we consider all the nearest neighbors
hoppings. The unit cell of the bilayer lattice structure contains
4 sites $A1$, $B1$, $A2$, and $B2$. The intra-layer hoppings
between the sites $A1-B1$ and $A2-B2$ are parameterized by the
single energy $t\approx 3 eV$. There are two type of inter-layer
hoppings: $A1-A2$ and $B1-B2$ which are characterized by the
energies $t_{\bot}\approx0.4 eV$ and $t_3\approx 0.3 eV$,
respectively.
\par
The resulting tight binding Hamiltonian is written in k-space. The
hexagonal first Brillouin zone of bilayer graphene contains 6
corners, among them two are inequivalent specifying the valleys
$K$ and $K'$. In the absence of inter-valley scattering
\cite{castro-rmp}, the valleys are degenerate and it is sufficient
to consider only one valley. Low energy excitations with 2D wave
vector ${\bf k}\equiv (k,q)$ around one of these valleys, say
valley $K$, is governed by Hamiltonian of the form \cite{falko},
\begin{eqnarray}
  {\cal H}(\bf k) =
\left(\matrix{
    0  & \hbar v k_- &  t_{\bot} & 0 \cr
    \hbar v k_+ & 0 & 0  & \hbar v_3 k_- \cr
    t_{\bot} & 0  & 0  & \hbar v k_+ \cr
    0 &\hbar v_3 k_+  & \hbar v k_- & 0}\right),
  \label{eq1}
\end{eqnarray}
which operates in the space of 4-component spinors of the form
$\Psi=e^{ikx+iqy}(\psi_{A1},\psi_{B1},\psi_{A2},\psi_{B2})$, where
each component determines the wave function amplitude at the
corresponding site of the bilayer unit cell. The characteristic
velocities $ v= 3ta/(2\hbar)$,and $v_3= 3t_3 a/(2\hbar)$ ($a$ is
the lattice constant) are associated with the hopping energies $t$
and $t_3$, respectively. We note that the complex wave vectors
$k_{\pm}=e^{\mp i\theta}(k\pm iq)$ depend on the misorientation
angle $\theta$.
\par
The quasiparticle spectrum $\varepsilon (k,q)$ is obtained from
the eigenvalue equation of Hamiltonian (\ref{eq1}), which
reads
\begin{eqnarray}
(\varepsilon^2-|\hbar v{\bf
k}|^2)^2-\varepsilon^2(t_{\bot}^2+ |\hbar v_3{\bf k}|^2)+|\hbar v_3{\bf k}|^2 t_{\bot}^2\nonumber\\
=(\hbar v)^2 t_{\bot} \hbar v_3(e^{3i\theta}(k-iq)^3
+e^{-3i\theta}(k+iq)^3). \label{eq2}
\end{eqnarray}
It is clear that the right hand side of this equation produces
trigonal warping of constant-energy lines, with an strength given
by the ratio $\delta=v_3/v$. Fig. \ref{fig1}(b) shows
constant-energy lines for three energies
$\varepsilon/\varepsilon_L=0, 0.8, 2$, where
$\varepsilon_L=t_{\bot}\delta^2/4$ is the characteristic energy at
which the Lifshitz transition takes place. While at $\varepsilon=
2 \varepsilon_L$ the energy line, despite its strong trigonal
warping, is a continuous line, at $\varepsilon= 0.8\varepsilon_L$
it breaks into four pockets whose enclosed area decreases with
lowering energy and finally at $\varepsilon=0$ shrinks into the
four points. The central point is located at $|{\bf k}|=0$ and
other three leg points at a constant distance $|{\bf k}|=\Delta
k=t_{\bot}v_{3}/(\hbar v^{2})$ from the center and in the
directions determined by angles
$\theta_n=\arctan(q_n/k_n)=\theta+2(n-1)\pi/3$, $n=1,2,3$ (see
Fig. \ref{fig1}(b)).
\par
Dirac point splitting will have two main effects on the transport
of carriers through the bilayer strip.  First it introduces a
length scale $\ell=\pi/\Delta k=(\pi\hbar
v)/(t_{\bot}\delta)\approx 50 nm$ as the effective range of the
scattering potential which could mix the states at the four Dirac
points. This length scale is an order of magnitude larger than
interlayer coupling length $l_{\bot}=\hbar v/t_{\bot}$ introduced
in Ref. \cite{beenakker07}. For realistic graphene samples of few
100 nm length, $L>\ell$ and the variations over this new length
scale should be taken into account. The potential profile varies
over the length $L$ of the strip.  For $L\lesssim\ell$ the
scattering potential is short enough to cause strong inter Dirac
point scattering and consequently the scattering states at these
points are mixed. On the other hand for $L\gg \ell$ the states of
the Dirac points are well separated and do not mix. Secondly, it
causes an anisotropy of the scattering states due to the
orientation of the leg points. As we will explain in the
following, these effects result in a length-dependent anisotropic
minimal conductivity for the bilayer strip.
\par

Within the scattering formalism we find the transmission amplitude
of electrons through the bilayer strip. An electronic state is
specified by the energy $\varepsilon$ and the transverse wave
vector $q$, which are conserved in the scattering process. We find
the eigenstates of Hamiltonian (\ref{eq1}) in three regions of
left ($x<0$) and right ($x>L$) electrodes and the bilayer strip
($0<x<L$). In general for a given $\varepsilon$ and $q$ there are
4 values of longitudinal momentum $k$ (solutions of Eq.
(\ref{eq2})) with 4 corresponding eigenstates in each region.
\par
Inside the strip at the neutrality point $\varepsilon=0$, the 4
solutions ($k_{i}$, $i=1,...,4$) of Eq. (\ref{eq2}) have
the form
\begin{eqnarray}
k_{1,2}=k^{\ast}_{3,4}=iq+\frac{\pi}{2\ell}e^{-3i\theta}(1\pm\sqrt{1+i\frac{8q\ell}{\pi}
e^{3i\theta}}).\label{eq3}
\end{eqnarray}
The corresponding eigenstates are given by
\begin{eqnarray}
\phi_{1,2}&=&e^{ik_{1,2}x+iqy}(0,-\frac{e^{-i\theta}}{l_{\bot}},k_{1,2}-iq,0),\label{eq4}\\
\phi_{3,4}&=&e^{ik_{3,4}x+iqy}(k_{3,4}+iq,0,0,-\frac{e^{i\theta}}{l_{\bot}}).\label{eq5}
\end{eqnarray}
In general for a given $q$ all 4 states inside the strip are
evanescent having complex $k_i (i=1,...,4)$ with exceptions of the
Dirac points with $q_0=0$ and $q_n=(\pi/\ell)\sin(\theta_n)$
($n=1,2,3$) at which two of $k_i$s are real representing
propagating states in the strip.
\par
Inside the highly doped electrode regions a very large potential
$-U_0$ is applied. We assume that for all states contributing in
transport $\hbar v q\ll U_0$ provided that $U_0$ be much larger
than all other energy scales in Hamiltonian (\ref{eq1}).
By this assumption the longitudinal momenta of all states inside
electrode regions have a constant magnitude $k_0=U_0/\hbar v\gg
q$. Inside each electrode for certain $q$ there are two
right-going eigenstates,
\begin{equation}
\phi^{R}_{\pm}=e^{ik_0 x+iqy}(1, e^{-i\theta},\pm 1,\pm
e^{i\theta}),
\label{eq6}
\end{equation}
and two left-going eigenstates,
\begin{equation}
\phi^{L}_{\pm}=e^{-ik_0 x+iqy}(1,-e^{-i\theta},\pm 1,\mp
e^{i\theta}).
\label{eq7}
\end{equation}
\par
For two left-going incident states from the left electrode ($x<0$)
the scattering states in three different regions have the form
\begin{equation}
\Psi_\pm=\left\{\begin{array}{ll} \raisebox{1mm}{$
\phi_{\pm}^{R}+r^{\pm}_{+}\phi_{+}^{L}+r^{\pm}_{-}\phi_{-}^{L} $}&
\raisebox{1mm}{$\hspace{3mm}x<0$,}\\
\sum_{i=1}^{4}c_{i}^{\pm}\phi_{i}&\hspace{5mm}0<x<L,\\
\raisebox{-1mm}{$t_{+}^{\pm}\phi_{+}^{R}+t^{\pm}_{-}\phi_{-}^{R}$}&\raisebox{-1mm}{$\hspace{3mm}x>L$.}\end{array}\right.
\label{eq8}
\end{equation}
where the coefficients $c_{i}^{\pm}$ and reflection and
transmission amplitudes $r^{\pm}_{\pm}$, $t^{\pm}_{\pm}$ have to
be determined by imposing the continuity condition of the wave
functions at the boundaries $x=0,L$.
\par
We calculate the conductance at zero temperature from
Landauer-Buttiker formula,
\begin{equation}
\frac{G}{G_0}= \frac{W}{2\pi}\int_{-\infty}^{\infty} T(q) dq,
\label{eq9}
\end{equation}
where
$T(q)=|t_{+}^{+}|^2+|t_{+}^{-}|^2+|t_{-}^{+}|^2+|t_{-}^{-}|^2$ is
the sum of transmission probabilities of the two states
$\Psi_{\pm}$, and $G_0=4 e^2/h$ is 4 times of conductance quantum
to take into account the valley and spin degeneracies. The
conductivity of the bilayer strip is obtained by the relation
$\sigma=(W/L)G$.
\par
We have obtained $T(q)$ as a function of the length $L$ and the
orientation angle $\theta$. For a short strip with $L\ll\ell$  the
transmission probability takes the form
\begin{equation}
T(q)=\frac{1}{\cosh^{2}[(q-q_c)L]}+\frac{1}{\cosh^{2}[(q+q_c)L]},
\label{eq10}
\end{equation}
which shows two maxima at the point $q=\pm q_c=\pm {\rm
arcsinh}(L/2l_{\bot})/L$. Around these points $T(q)$ decays
exponentially within a scale of order $\Delta q\sim 1/L$ which for
a short strip is much larger than $\Delta k=\pi/\ell$. Thus $T(q)$
is almost constant within the scale $\Delta k$ which implies that
the Dirac point splitting and the associated anisotropy of the
spectrum are not revealed in the transmission of the carriers.
This is the result of strong mixing of the states around the 4
Diract points via scattering through the bilayer strip. The
resulting isotropic minimum conductivity $\sigma^{min}_{\bot}=2
\sigma^{min}_{0} $ which is twice of the minimum conductivity for
a single layer.

\begin{figure}
\begin{centering}\includegraphics[width=8cm]{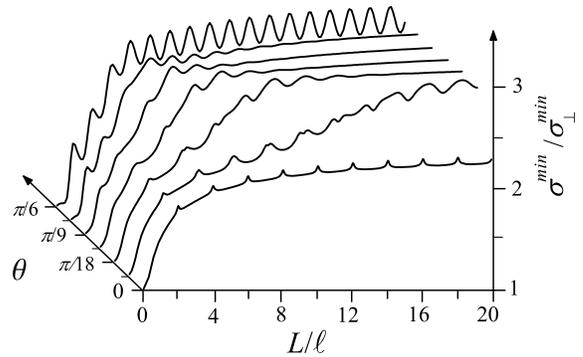}\par\end{centering}
\caption{\label{fig2}{\small Minimal conductivity in units of
$\sigma^{min}_{\bot}=(8/\pi)(e^2/h)$ of a wide strip of undoped
bilayer graphene versus its length $L$ for different orientations
$\theta$ of the hexagonal lattice symmetry axis with respect to
the electrodes. $L$ is measured in units of $\ell=\pi/\Delta k$
with $\Delta k$ being the k-space distance of two trigonally split
Dirac points. The isotropic minimal conductivity
$\sigma^{min}_{\bot}$ at $L\ll \ell$ increases with $L/\ell$ to
higher values and becomes anisotropic.}}
\end{figure}

\par
For a finite $L/\ell$ the minimum conductivity increases above
$\sigma^{min}_{\bot}$. This is shown in Fig. \ref{fig2} where we
have plotted $\sigma^{min}$ as a function of $L/\ell$ for
different orientations $\theta$.  For $L/\ell\gtrsim 1$ the
increased minimum conductivity finds a $\theta$-dependence as the
result of trigonal splitting of the Dirac point. In this case the
anisotropy of $\sigma^{min}$ is spread over the range
$0<\theta<\pi/6$. We note that the increase in $\sigma^{min}$ is
associated with an oscillatory variation due to quantum
interference effects in bilayer strip. Increasing $L/\ell$ further
to approach the limit of a long strip $L\gg \ell$ the range of
anisotropy becomes narrower. In this limit and for the angles not
too close to $\theta=0,\pi/6$  we obtain the following relation
for the transmission probability
\begin{equation}
T(q)=\sum_{n=0,3}\frac{1}{\cosh^2[\alpha_n(\theta)(q-q_n(\theta))L]},
\label{eq11}
\end{equation}
where the summation is taken over 4 transverse coordinates of the
Dirac points, $q_n$, with $\alpha_0=1$ and
$\alpha_n(\theta)=3/(5-4\cos[2\theta_n])$ for $n=1,2,3$. This
result shows that the transmission probability consists of 4
resonant peaks at the points $q_n$, whose width is of order
$\Delta q \sim 1/L<<\Delta k$. The effect of the Dirac point
splitting is, thus, revealed in the transmission process. For
$\theta$ approaching $0$ and $\pi/6$ the transverse coordinates of
the two resonant peaks $q_0\rightarrow q_1$ and $q_1\rightarrow
q_2$, respectively, and the corresponding peaks overlap. For these
cases Eq. (\ref{eq11}) is not applicable, since we have considered
four Dirac points contribution independently in this equation. In
the case of $\theta=0$ from Eqs. (\ref{eq8}) and (\ref{eq9}) we
find that the contribution of the conductivity from the peaks at
$q_0=q_1=0$ is of order $\delta^2 e^2/h$ which is negligibly
small. So there are only contributions from the points $q_2,q_3$
which results in a minimum conductivity
$\sigma^{min}_{\theta=0}=(7/3)\sigma^{min}_{\bot}$. In contrast to
$\theta=0$ case, for $\theta=\pi/6$ the overlapped resonant peaks
makes the same contributions as two independent peaks. This gives
a minimum conductivity $\sigma^{min}=3\sigma^{min}_{\bot}$. We
find that for a long strip this value of the minimum conductivity
is valid for all orientations $0\lesssim \theta < \pi/6$ except of
the angles $\theta \lesssim \ell/L$ where the two peaks $q_0, q_1$
have a significant overlap. Thus the minimum conductivity is
anisotropic over the range $\Delta\theta\sim \ell/L$ and an
amplitude $\Delta \sigma^{min}=(2/3)\sigma^{min}_{\bot}$.

\par
In conclusion, we have studied conductivity of a wide strip of
undoped bilayer graphene which connects two highly doped electrode
regions. We have shown that due to the trigonal splitting of Dirac
point at zero Fermi energy, the minimal conductivity
$\sigma^{min}$ of the strip finds a dependence on the lattice
symmetry axis orientation $\theta$ with respect to the electrodes.
The anisotropy of $\sigma^{min}$ depends on the length of strip
$L$ as compared to the characteristic length scale
$\ell=\pi/\Delta k\simeq 50 nm$ determined by the inverse of the
k-space separation of two Dirac points. For a short strip of $L\ll
\ell$, $\sigma^{min}$ takes an isotropic universal value
$\sigma^{min}_{\bot}=(8/\pi)e^2/h$. For longer strips the minimal
conductivity increases above this value in an anisotropic way. We
have found that in the limit of $L\gg \ell$ the anisotropic
minimal conductivity grows from $(7/3)\sigma^{min}_{\bot}$ at
$\theta=0$ to $3\sigma^{min}_{\bot}$ when orientation is changed
by the angle $\Delta \theta\sim \ell/L$.

\end{document}